\documentclass[12pt,preprint]{aastex}
\usepackage{epsfig}
\usepackage{psfig}
\usepackage[usenames,dvips]{color}
\def\msun{M_{\odot}}

\shorttitle{{Diffuse X-ray emission in globular cluster cores}}
\shortauthors{Hui et al.}
\begin{document}

\title{Diffuse X-ray emission in globular cluster cores}

\author{
C. Y. Hui\altaffilmark{1}, K. S. Cheng\altaffilmark{2}, and Ronald
E. Taam\altaffilmark{3, 4, 5} }

\altaffiltext{1} {Max-Planck Institut f\"ur Extraterrestrische
Physik, 85741 Garching bei M\"unchen, Germany} \altaffiltext{2}
{Department of Physics, University of Hong Kong, Pokfulam Road, Hong
Kong} \altaffiltext{3} {Department of Physics and Astronomy,
Northwestern University, 2131 Tech Drive, Evanston, IL 60208}
\altaffiltext{4} {Academia Sinica Institute of Astronomy and
Astrophysics - TIARA, P.O. Box 23-141, Taipei, 10617 Taiwan}
\altaffiltext{5} {Academia Sinica Institute of Astronomy and
Astrophysics/National Tsing Hua University - TIARA, Hsinchu, Taiwan}

\begin{abstract}
The unresolved X-ray emission in the cores of 10 globular clusters hosting
millisecond pulsars is investigated. Subtraction of the known resolved point
sources leads to detectable levels of unresolved emission in the core region
of M28, NGC 6440, M62, and NGC 6752. The X-ray luminosities in the 0.3-8 keV
energy band of this emission component were found to lie in the range $\sim
1.5 \times 10^{31}$ ergs s$^{-1}$ (NGC 6752) to $\sim 2.2 \times 10^{32}$
ergs s$^{-1}$ (M28). The lowest limiting luminosity for X-ray source detections
amongst these four clusters was $1.1 \times 10^{30}$ ergs s$^{-1}$ for NGC 6752.
The spectrum of the unresolved emission can be fit equally well by a power-law,
a thermal bremsstrahlung model, a black body plus power-law, or a 
thermal bremsstrahlung model plus black body component.
The unresolved emission is considered to arise from the cumulative contribution
of active binaries, cataclysmic variables, and faint millisecond pulsars 
with their associated pulsar wind nebulae.  
In examining the available X-ray data, no evidence for any pulsar wind 
nebular emission in globular clusters is found. 
It is shown that the X-ray luminosity
contribution of a faint source population based on an extrapolation of the luminosity
function of detected point sources is compatible with the unresolved X-ray emission 
in the cores of NGC 6440 and NGC 6752.  Adopting the same slope for the luminosity 
function for M62 as for NGC 6440 and NGC 6752 leads to a similar result for M62. 
For M28, the contribution from faint sources in the core can attain a level comparable with the 
observed value if a steeper slope is adopted. 
The characteristics on the 
faint source population as constrained by the properties of the unresolved X-ray 
emission are briefly discussed.
\end{abstract}


\section{INTRODUCTION \label{intro}}

The point X-ray source population in globular cluster stellar systems has received
much attention in recent years as a direct result of the high angular resolution
studies afforded by the Chandra X-ray Observatory.  The observational studies have
provided an invaluable probe for investigating the binary star population containing
compact objects in these systems. For example, the globular clusters with X-ray sources
brighter than $4 \times 10^{30}$ ergs s$^{-1}$ have revealed a strong correlation between
the number of such sources with the normalized encounter rate in the cluster (see Pooley
et al. 2003) with implications on their formation by dynamical means and on the
equilibrium state of the globular cluster systems themselves (Fregeau 2008).

In addition to the studies of the bright X-ray sources, such as low mass X-ray binaries,
investigations of the fainter X-ray source population have also revealed important diagnostic
information on the importance of dynamical interactions in forming cataclysmic variable
stars (Pooley \& Hut 2006). Furthermore, the recognition that recycled millisecond pulsars
contribute to this faint population (Grindlay et al. 2001) poses important questions
regarding their formation, evolution, and retention within the dense stellar environment of
globular clusters. In this regard, we note that the observations in the radio wavelength
region have been particularly important in characterizing this neutron star subpopulation,
with over 140 such radio pulsars now known to be present in globular clusters.

Progress in the theoretical understanding of the formation of compact objects in binary
systems within globular clusters has advanced in parallel based on phenemological studies
of the importance of dynamical processes in the formation of cataclysmic variables (e.g.,
Pooley \& Hut 2006) and binary population synthesis studies which include stellar dynamical
interactions in a model globular cluster system (Ivanova et al. 2007, 2008). An important
finding is the realization that the retention of neutron stars in globular clusters may
result from the formation of neutron stars via core collapse induced by an electron capture
process, confirming an earlier suggestion by Podsiadlowski et al. (2004).

In contrast to the point source population, little work has been directed toward the study
of the unresolved faint point source population or diffuse X-ray emission
in globular cluster systems.  Recently, Okada et al. (2007) have investigated the extended
emission in 6 globular clusters, finding evidence for the presence of diffuse emission
in 47 Tuc and NGC 6752 offset from the cluster center.  The origin of this emission was
suggested as resulting from the production of shocks associated with the motion of the
globular cluster through the Galactic halo. In this study, we focus on the unresolved or
diffuse emission in the high density core regions of globular clusters where the presence
of millisecond pulsars (MSPs) suggests that dynamical interactions were important in the
formation of these objects as well as of other compact binary systems. The possible
existence of such emission, therefore, may reflect a consequential outcome of the formation (and
perhaps destruction) of binaries in cluster cores.  To assess the contribution of
faint X-ray sources such as active binaries, cataclysmic variables, and faint MSPs with 
their pulsar wind nebulae (PWNe) to the unresolved X-ray emission in the core
region, we have carried out an observational investigation to detect the diffuse X-ray
emission component in the core region of select globular clusters. In the next section,
we present the observations and data analysis of a number of globular clusters which have
been studied by the Chandra X-ray Observatory with sub-arcsecond angular resolution so that
the diffuse emission can be distinguished from the X-ray point source contribution.
Based on the spectral characteristics and luminosity of the unresolved emission
in the core region, we examine the potential contribution of unresolved point sources and the
diffuse emission of PWNe from MSPs in \S 3.  The
implications of our results and suggested follow up observations are discussed in the
last section.

\section{X-RAY OBSERVATIONS AND DATA ANALYSIS \label{obs}}

In order to better isolate the X-ray point sources from the
unresolved emission in the cluster cores, our analysis was limited
only to the data obtained by the Chandra X-ray Observatory which
provides sub-arcsecond angular resolution.  Specifically, we have
undertaken a systematic search of the Chandra data archive for
observations of all globular clusters that belong to the group that
host MSPs. Since we are only interested in imaging spectroscopic
data, only the data obtained by the Advanced CCD Imaging
Spectrometer (ACIS) without any grating was selected. As the regions
of interest in our analysis are the cluster cores, only clusters
having core radii $r_{c}\gtrsim 5"$, which are sufficiently large
for our study, were chosen.  Given these selection criteria, 10
globular clusters were selected. The corresponding observational
details are summarized in Table~\ref{obs_info} and the physical
properties of these clusters are summarized in Table~\ref{gc_info}.

We have deliberately excluded 47~Tucanae and Terzan~5 in this work because the densities of
X-ray point sources are very high in these two clusters (cf. Fig.~1 in Heinke et al. 2006 and
Fig.~2 in Heinke et al. 2005).  The removal of all known point sources in these clusters
resulted in insufficient photons for any detailed analysis.

In our study, standard processed level-2 data were used on the cluster cores (i.e. a circular
region with radius of $r_{c}$).
All the detected point sources were removed within $r_{c}$ from the data. We simulated the model
point spread function (PSF) at 1~keV for the point sources by using MARX.
For the inputs of the simulations, we adopted the same roll, off-axis angle, and determined
the density of rays by using the effective area of the CCD and the corresponding detected
counts for each source.  With the consideration of the PSF wings, the excluded regions
were optimized so as to mininize the
contamination from these discrete X-ray sources in order to retain as many unresolved X-rays
in the core region as possible at the same time.

Since our regions of interest were all
located on the back-illuminated ACIS-S3 CCD chip which has a superior low energy response,
we performed our analysis in the energy band of $0.3-8$ keV.
The response files were computed by using the tools MKRMF and MKARF
in CIAO 3.4.  Ultilizing the updated calibration data, CALDB 3.4.1,
these files incorporate the corrected degradation for the quantum
efficiency in the ACIS CCD.  All the spectral fittings were
performed with XSPEC 12.3.1, and the quoted errors of the
spectral parameters are $1\sigma$ for 1 parameter of interest.

\subsection{M28 (NGC~6626)}

M28 has been observed by Chandra on five occasions. Three were performed with the Advanced
CCD Imaging Spectrometer (ACIS-S) and two were carried out with the High Resolution Camera
(HRC-S). Since the HRC-S data do not provide any spectral information, we do not consider
these datasets in this paper. The ACIS-S observations were scheduled on 4 July 2002, 4 August
2002 and 9 September 2002 with an effective exposure of $\sim13-14$ ks (Table~\ref{obs_info}).
Using these datasets, Becker et al. (2003) detected 46 discrete X-ray point sources in total
by means of a wavelet detection algorithm. We have re-run the source detection on the merged
data and independently confirmed the results reported by Becker et al. (2003). On the basis of
the detection limit, a limiting luminosity at a level of $\sim 3.5\times10^{30}$ ergs s$^{-1}$
is inferred (see the last column in Table~\ref{gc_info}).
The emission properties of these point sources have been characterized by Becker et al. (2003)
and Becker \& Hui (2007), and include the X-ray counterparts of MSPs M28A and M28H.

With a view to obtain better statistics for the analysis, all ACIS
observations were combined to produce improved images.  Prior to the
merging process, the aspect offsets for each observation, which is a
function of the spacecraft roll angle, have been carefully checked
and corrected. All three datasets were then merged to a common
position. After the subtraction of all known point sources, we estimated that the
total contribution of the PSF wings centered on these sources in the remaining photons should
be $\sim10\%$.  The contamination is dominated by the pulsar M28A and the brightest source in the field
(source \#26 in Tab.~3 of Becker et al. 2003), which was identified as a quiescent low mass X-ray
binary candidate.

We also estimated the possible contribution of background sources within $r_{c}$. By fitting the
projected surface density of M28, Becker et al. (2003) obtained the best-fitted value of
$0.36\pm0.22$ background sources per arcmin$^{2}$.  This is comparable with the estimate from
the Chandra Deep Field South (Rosati et al. 2002). Assuming the background sources are uniformly
distributed, we conclude that the contribution of background sources in our region of interest is
negligible.

\begin{figure}
\begin{center}
\psfig{figure=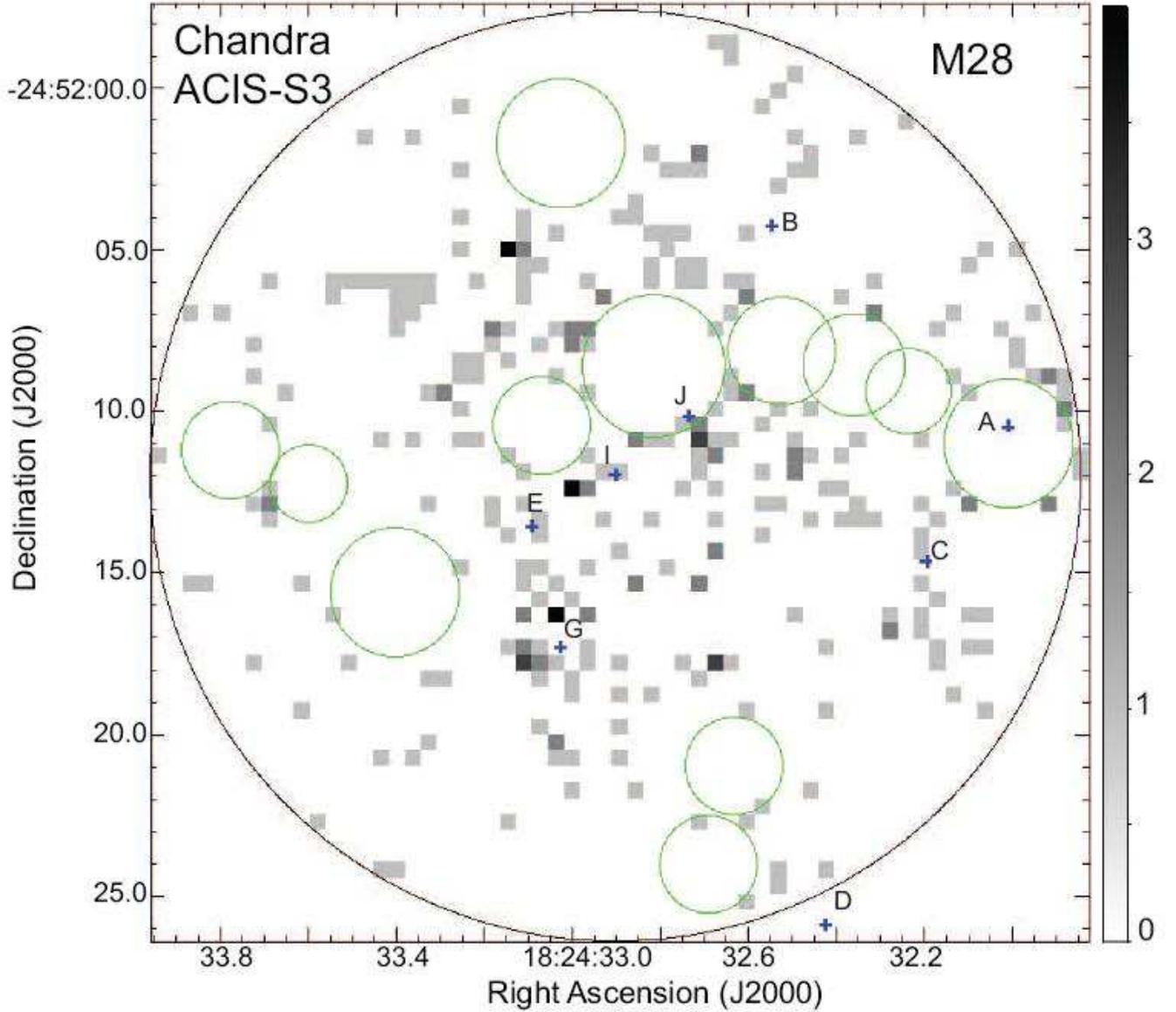,width=18cm,clip=}
\end{center}
\vspace{-1cm}
\caption[]{The raw X-image of the unresolved emission within the core radius of
M28 as observed by Chandra ACIS-S3 CCD in $0.3-8$ keV. The excluded regions of
the detected point sources are illustrated as green circles. The black circle indicates 
the region of 1 core radius centered at the optical center. The image is created by
merging all three ACIS observations. The radio timing positions of eight MSPs in the
field-of-view of this image are illustrated by the crosses. The count scale in each pixel 
is illustrated by the grey-scale bar.}
\label{m28_coord}
\end{figure}

The raw X-ray image of the unresolved emission from the cluster core (i.e. a circular region of
interest with $r=r_c$) is shown in Figure~\ref{m28_coord}. The image is binned with a factor
of 0.5 arcsec.  The radio timing positions of eight MSPs in this field of view are illustrated
as crosses in Figure~\ref{m28_coord} (cf. Tab.~2 in Becker \& Hui 2007). We notice that two clumps
of photons are close to the timing positions of MSPs M28G and M28J. While the feature near M28J is
adjacent to the subtracted region of a detected source and thus can possibly be a result arising 
from the imperfect subtraction, the clump close to M28G is likely to be a faint source just below 
the detection limit of resolved sources. 

\begin{figure}
\psfig{figure=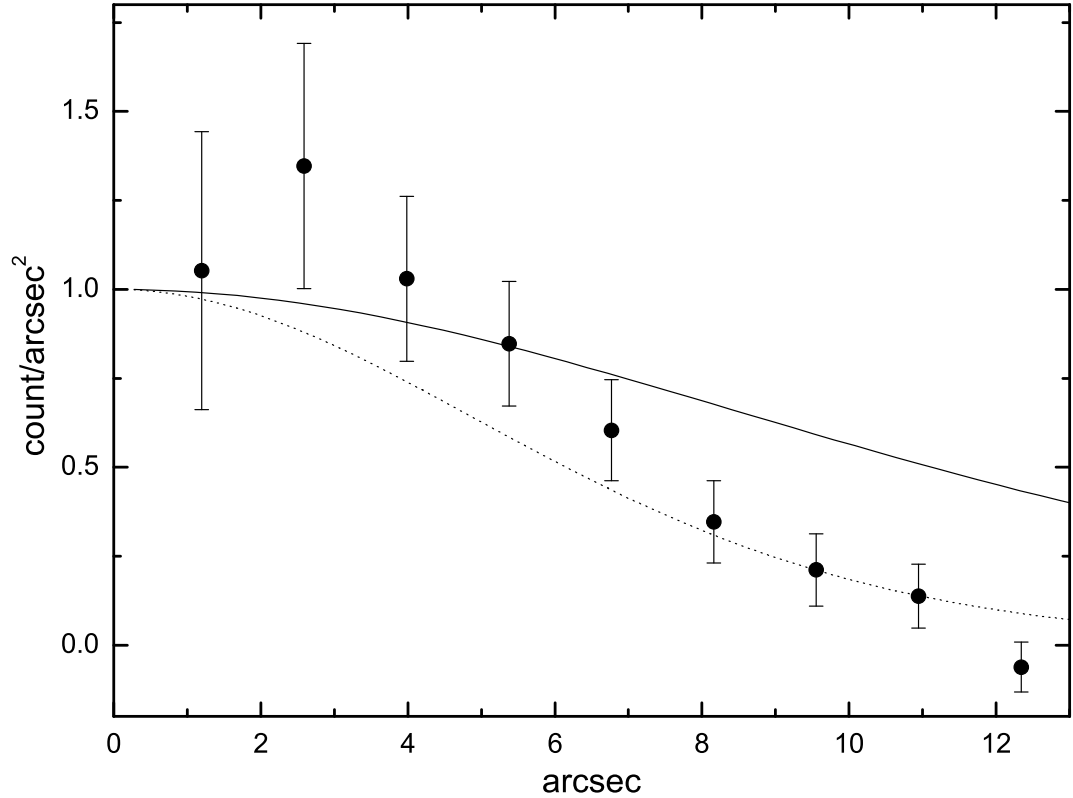,width=16cm,clip=}
\caption[]{The radial brightness profile of the unresolved emission within the core radius of
M28 as observed by Chandra ACIS-S3 CCD in $0.3-8$ keV. The solid line and the dotted line 
(with the vertical quantity in arbitrary units) represent the
best-fit projected surface density profile of the detected point sources and the profile with the
parameters stretched to $1\sigma$ uncertainties reported by Becker et al. (2003) respectively.} 
\label{m28_sur_bri}
\end{figure}

We have also computed the surface brightness profile of the unresolved emission in M28
which is displayed in Figure~\ref{m28_sur_bri}.  Comparing this profile with the best-fit 
projected surface density $S(r)$ of the detected X-ray point sources in the cluster (cf. 
Becker et al. 2003), we notice that the best-fit $S(r)$ (solid line in Fig.\ref{m28_sur_bri})
falls much slower than the brightness profile of the unresolved emission. Within $1\sigma$ 
errors of the parameters in $S(r)$ reported by Becker et al. (2003) (dotted line in Fig.\ref{m28_sur_bri}), 
the radial profile of the unresolved emission has a similar trend as that of the detected point sources. 
A comparison with radial distributions of different populations of X-ray emitting 
objects (e.g. white dwarfs and MSPs) may provide insights into the nature of the unresolved 
X-ray emission.

The spectrum of the diffuse emission in all datasets was extracted from a circle of $r=r_{c}$
centered on the cluster center. In addition, the background spectrum was extracted from a source
free region within a 10 arcsec radius centered at RA=$18^{\rm h}24^{\rm m}32.209^{\rm s}$,
Dec=$-24^{\circ}51'46.11"$ (J2000).  After background subtraction, there are totally $259\pm16$ net
counts available for the spectral analysis.  The background contribution was estimated to be $\sim15\%$.
The net counting rates for the observations in July, August and September are $(6.4\pm0.8)\times10^{-3}$
cts/s, $(5.2\pm0.8)\times10^{-3}$ cts/s, and $(6.4\pm0.9)\times10^{-3}$ cts/s.  Each spectrum was
dynamically binned so as to have at least 10 counts per bin.  To better constrain the spectral
properties, we fitted the spectra obtained from all observations simultaneously. Four different
spectral models were adopted, corresponding to a power-law, thermal bremsstrahlung, power-law plus
black body, and thermal bremsstrahlung plus black body.  The results are summarized
in Table~\ref{gc_spec}.

\begin{figure}
\centerline{\psfig{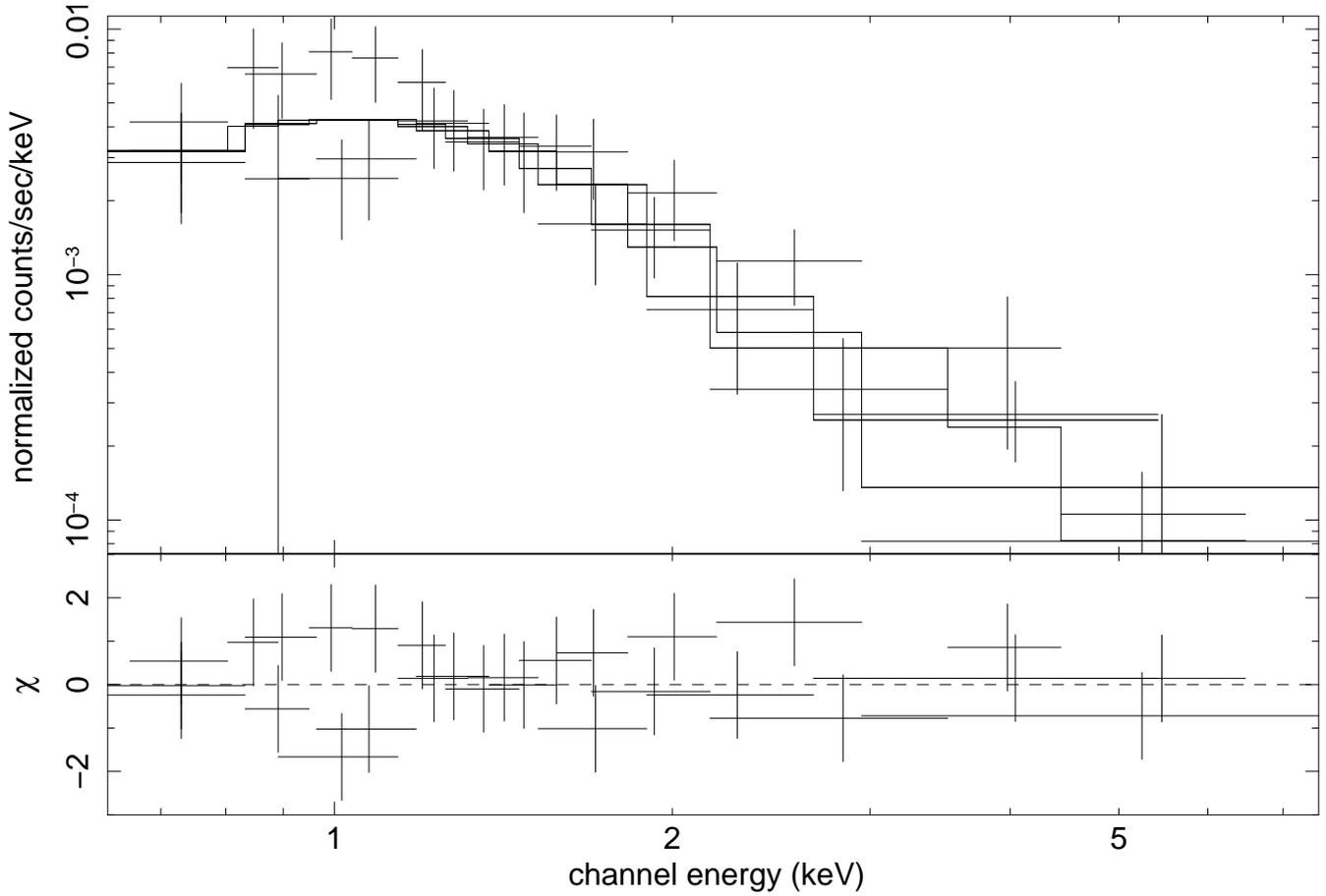}}
\caption[]{Energy spectrum of the unresolved X-rays in the core of M28, as observed with
ACIS-S3 detectors and fitted to an absorbed power-law model
({\it upper panel}) and contribution to the $\chi^{2}$ fit statistic
({\it lower panel}).}
\label{m28_spec}
\end{figure}

We found that a power-law model fits the data reasonably well
($\chi^{2}_{\nu}=0.83$ for 25 D.O.F.). This model yields a column
density of $n_{H}=2.75^{+0.78}_{-0.84}\times10^{21}$ cm$^{-2}$ and a
photon index of $\Gamma=2.35^{+0.41}_{-0.25}$.
The energy spectrum is displayed in Figure~\ref{m28_spec}. No systematic 
deviation of the fitting residual is found for the absorbed power-law model in $0.3-8$ keV 
(see lower panel of Fig.~\ref{m28_spec}).
At a distance of 5.5 kpc, the unabsorbed luminosity deduced from this model is
$L_{X}=2.54^{+1.27}_{-0.62}\times10^{32}$ erg s$^{-1}$ in the energy range of $0.3-8$ keV.
In addition to the power-law model, we found that the thermal bremssstrahlung model
can describe the data equally well ($\chi^{2}_{\nu}=0.82$ for 25 D.O.F.) with
$n_{H}=1.50^{+0.90}_{-0.76}\times10^{21}$ cm$^{-2}$ and a temperature
of $kT=2.58^{+1.09}_{-0.75}$~keV. This best-fit model implies an absorption-corrected
luminosity of $L_{X}=1.67^{+0.99}_{-0.62}\times10^{32}$ erg s$^{-1}$ at 5.5 kpc.
The best-fit values of $n_{H}$ in both these models are consistent
with that deduced from the foreground reddening (i.e. $\sim2.2\times10^{21}$ cm$^{-2}$).

Since the X-ray emission from most of the MSPs detected in 47~Tuc is thermally dominant
(Bogdanov et al. 2006), it is instructive to examine the possible contribution of a MSP
population by including a black body component despite the fact that the single component
of power-law/thermal bremsstrahlung model can well describe the data already. To better
constrain the contribution of the black body component, we fixed $n_{H}$ at the value
deduced from the optical reddening and the black body temperature at $2.1\times10^{6}~K$
which is the mean value inferred from the MSP population in 47~Tuc (Bogdanov et al. 2006).
Acceptable fits were found, resulting in a black body contribution of $\lesssim9\%$
and $\sim12\%$ in the power-law plus black body (PL+BB) and the thermal bremsstrahlung
plus black body (BREMSS+BB) fits respectively. 

The robustness of the best-fitted parameters quoted in this paper
were checked by repeating all the aforementioned spectral fitting by
incorporating the background spectrum extracted from different
source-free regions in the observed data as well as the ACIS ``blank
sky" background files. It is found that within $1\sigma$ errors the
spectral parameters inferred from independent fittings are all
consistent with each other.

\subsection{NGC~6440}

NGC~6440 has been observed by the Chandra X-ray Observatory
on three occasions (see Table~\ref{obs_info}).  All the
observations were carried out by ACIS-S and scheduled on 4 July 2000, 13 August 2001 and 27 June
2003 respectively. Using the data obtained from the observation on 4 July 2000, Pooley et al.
(2002) detected 24 discrete X-ray point sources to a limiting luminosity of $\sim2\times10^{31}$
ergs~s$^{-1}$ within the half-mass radius (cf. Table~2 in Pooley et al. 2002), 16 of which
are located inside the core radius. The source CX~1 as listed in the Table~2
of Pooley et al. (2002) has been identified as an X-ray transient (in 't Zand et al. 2001).
While NGC~6440 was observed on 4 July 2000 when CX~1 was quiescent, a target-of-opportunity (TOO)
observation on 13 August 2001 detected an outburst from
a transient with a luminosity of $\sim9\times10^{35}$ ergs~s$^{-1}$ (in 't Zand et al.
2001).  Owing to the outburst, the data obtained from the TOO observation were seriously
compromised by pileup effects.  Therefore, we do not include this dataset in the subsequent analysis.

The observations on 4 July 2000 and 27 June 2003 have an effective exposure of
$\sim 23$ ks and $\sim24$ ks respectively. Aspect offsets have been carefully
checked and corrected for each dataset. As our primary interest is the unresolved X-rays
in the cluster core, we first removed all the 16 resolved point sources within $r_{c}$
from the data. After excluding the contributions from the point sources as reported
in Table~2 in Pooley et al. (2002), we compared the count rates of the residual emission
within $r_{c}$ from both observations. We noticed that the net count rate in the 2003
observation is higher than that of the 2000 observation at a level of $\sim3\sigma$.

This led to a closer inspection of both datasets, and images were produced with a
bin-size of 0.5 arcsec from both observations. Applying the wavelet source detection algorithm
on each image individually, an additional source, besides the point sources reported by Pooley
et al. (2002), was detected close to the cluster center in the 2003 observation.  This source
is located at RA=$17^{\rm h}48^{\rm m}52.656^{\rm s}$, Dec=$-20^{\circ}21^{'}34.25^{"}$
(J2000). On the other hand, with the same detection algorithm, the source was not detected
in the 2000 observation. This led to the detection of a new transient X-ray source in
NGC~6440. The background subtracted count rates of the X-ray transient are
$(2.07\pm 0.96)\times10^{-4}$ cts/s and $(1.03\pm 0.21)\times10^{-3}$ cts/s in
2000 observation and 2003 observation respectively. The significance of the difference in
count rates is at a level of $>3\sigma$.

Since a new source is detected in the core of NGC~6440, in addition to the sources reported by
Pooley et al. (2002), we further subtracted its contribution in both datasets. After removing
all the known point sources, the total contribution of the PSF wings centered
on these point sources in the remaining photons within the core was estimated to be $\sim18\%$.
The net count rates of the unresolved X-ray emission within $r_{c}$ are $(1.15\pm0.24)\times10^{-3}$
cts/s and $(1.13\pm0.24)\times10^{-3}$ cts/s for the 2000 observation and 2003 observations
respectively. Thus, the count rates from both observations are fully consistent with each other,
and the lack of variability justifies combining both datasets to improve the photon statistics.

We extracted the spectrum of the unresolved X-ray emission in both datasets
from a circle of $r=r_{c}$ centered on the cluster center.
The background spectrum was extracted from a source free region
within a 10 arcsec radius centered at RA=$17^{\rm h}48^{\rm m}54.324^{\rm s}$,
Dec=$-20^{\circ}21'25.80"$ (J2000).  After background subtraction, there
are totally $54\pm7$ net counts available for the spectral analysis.
The background contribution was estimated to be $\sim11\%$.
Each spectrum was dynamically binned so as to have at least 10 counts per bin.
We fitted both spectra simultaneously to better constrain the spectral properties.

In view of the limited photon statistics, we fixed the column
density at the reddening inferred value (i.e.
$n_{H}=5.9\times10^{21}$ cm$^{-2}$) for all the spectral fits.
A power-law model can describe
the data fairly well ($\chi^{2}_{\nu}=0.39$ for 4 D.O.F.)
which yields a photon index of $\Gamma=2.04^{+0.50}_{-0.46}$ and
an unabsorbed luminosity of $L_{X}=2.00^{+1.71}_{-0.76}\times10^{32}$
erg s$^{-1}$ in the $0.3-8$ keV energy band. A
thermal bremsstrahlung model can also describe the data
($\chi^{2}_{\nu}=0.44$ for 4 D.O.F.), yielding a temperature of
$3.06^{+6.18}_{-1.25}$~keV and an unabsorbed luminosity of
$L_{X}=1.57^{+1.51}_{-0.74}\times10^{32}$ erg s$^{-1}$.

Following the procedure described in \S2.1, we have also estimated the
possible contribution of a faint MSP population in NGC~6440
by incorporating a black body component in the spectral fits. This spectral
component is found to contribute $\sim17\%$ and $\sim7\%$ in the PL+BB and
BREMSS+BB fits respectively.

\subsection{M62 (NGC~6266)}

M62 was observed on 12 May 2002 with ACIS-S. The effective exposure of
this observation was $\sim 62$ ksec. Using this data, Pooley et al. (2003) detected
51 X-ray point sources within a half-mass radius of $1'.23$. However, the
details of these X-ray sources have not been reported in the literature. We have
independently performed the source detection. Using a wavelet detection algorithm, we have
detected 12 discrete sources within $r_{c}$ at a detection limit of $\sim7\times10^{30}$
erg~s$^{-1}$.

After the removal of all detected point sources in the core, we estimated that
the total contribution of the PSF wings centered on these point sources in the remaining
photons is about 20\%.

The spectrum of the unresolved X-ray emission in both datasets was extracted from a circle
of $r=r_{c}$ centered on the cluster center. The background spectrum was extracted from a
source free region within a 10 arcsec radius centered at RA=$17^{\rm h}01^{\rm m}10.610^{\rm s}$,
Dec=$-30^{\circ}06'27.63"$ (J2000).  After background subtraction, $164\pm13$ net counts
were available for the spectral analysis. The background contribution is rather small which
was estimated to be $\sim5\%$.  The spectrum was dynamically binned so as to have at least
15 counts per bin.

A power-law model can describe the data very well ($\chi^{2}_{\nu}=0.43$ for 8 D.O.F.).
In this case, the best-fitted model yields a hydrogen column density of $n_{H}<5.19
\times10^{20}$ cm$^{-2}$, a photon index of $\Gamma=1.54^{+0.22}_{-0.16}$ and an unabsorbed
luminosity of $L_{X}=1.50^{+0.62}_{-0.36}\times10^{32}$ erg s$^{-1}$ in the $0.3-8$ keV
energy band.  The inferred $n_{H}$ is small in
comparison with the value inferred from the average reddening (i.e. $n_{H}=2.6\times10^{21}$
cm$^{-2}$). However, this does not invalidate the spectral results reported here as the
reddening across the cluster is found to be highly varied (Possenti et al.
2003; Minniti, Coyne, \& Claria 1992).  We have also attempted to fix
$n_{H}$ at the value inferred by the average reddening, however, the power-law model was
not found to describe the data below 1 keV and above 3 keV, resulting in a worse goodness-of-fit
($\chi^{2}_{\nu}=1.03$ for 9 D.O.F.).

The thermal bremsstrahlung model can also describe the data with a comparable goodness-of-fit
($\chi^{2}_{\nu}=0.58$ for 8 D.O.F.). This best-fit model also yields a rather low
column density of $n_{H}<3.48\times10^{20}$ cm$^{-2}$ with a temperature of $kT=9.23^{+17.50}_{-4.10}$
keV and an unabsorbed luminosity of $L_{X}=1.41^{+0.26}_{-0.27}\times10^{32}$ erg s$^{-1}$ in
the $0.3-8$ keV energy band.

In the PL+BB and BREMSS+BB fits, the black body component with the temperature fixed at $2.1\times10^{6}~K$
is found to contribute $\sim35\%$ and $\sim32\%$ respectively.

\subsection{NGC~6752}

NGC~6752 was observed by the Chandra X-ray Observatory on two occasions (see
Table~\ref{obs_info}). All the observations were carried out by
ACIS-S and scheduled on 15 May 2000, and 10 February 2006. Using the
data obtained from the observation on 15 May
2000, Pooley et al. (2002b) detected 19 discrete X-ray point
sources within the half-mass radius (cf. Table~1 in Pooley et al.
2002b). We have re-analysed the X-ray image of the cluster by
combining both observations and inferred
a limiting luminosity of $1.1\times10^{30}$ ergs~s$^{-1}$.
Nine of the resolved X-ray sources are located inside the core radius.
After removing these point sources, we estimate that the total contribution of the PSF
wings centered on these sources in the remaining unresolved photons
is about $25\%$.

We extracted the spectrum of the unresolved X-ray emission in both
datasets from a circle of $r=r_{c}$ centered on the cluster center.
The background spectrum was extracted from a source free region
within a 10 arcsec radius centered at RA=$19^{\rm h}10^{\rm
m}47.967^{\rm s}$, Dec=$-59^{\circ}59'05.36"$ (J2000). The
background subtracted count rates for the observations in 2000 and
2006 are $(1.06\pm0.23)\times10^{-3}$ cts/s and
$(0.95\pm0.20)\times10^{-3}$ cts/s. The count rates from both
observations are thus consistent with each other within $1\sigma$
and justifies combining both datasets to improve the photon
statistics. After background subtraction,  $67\pm8$ net counts were
available for the spectral analysis.  The background contribution
was estimated to be $\sim26\%$.  The spectrum was dynamically binned
so as to have at least 10 counts per bin.

In view of the limited photon statistics, we fixed the column
density at the reddening inferred value, i.e.
$n_{H}=2.2\times10^{20}$ cm$^{-2}$. A power-law model can provide a
reasonable description of the data ($\chi^{2}_{\nu}=0.95$ for 7
D.O.F.). The best-fit model yields a photon
index of $\Gamma=1.94^{+0.39}_{-0.36}$ and an
unabsorbed luminosity of
$L_{X}=1.50^{+0.75}_{-0.42}\times10^{31}$ erg s$^{-1}$ in
$0.3-8$ keV. Fitting the data with a thermal bremsstrahlung
model also results in a good description ($\chi^{2}_{\nu}=0.89$ for
7 D.O.F.) resulting in a temperature of $2.38^{+3.06}_{-1.00}$~keV
and an unabsorbed luminosity of $L_{X}=1.19^{+1.12}_{-0.56}\times10^{31}$
erg s$^{-1}$. In the PL+BB and BREMSS+BB fits, the black body component
with the temperature fixed at $2.1\times10^{6}~K$
is found to contribute $\lesssim15\%$ and $\lesssim28\%$ respectively.

\subsection{Other clusters}

For the other selected globular clusters, no conclusive evidence for
an excess of unresolved X-ray emission in their cores is found.
After removing all the resolved X-ray point sources, we found that
the background contributions in the remaining photons within these
cluster cores are $\gtrsim65\%$.

\section{NATURE OF THE UNRESOLVED CORE X-RAY EMISSION \label{nature}}

To probe the physical nature of the unresolved X-ray emission in the
cores of globular clusters, there are two possibilities to consider.
Namely, the emission is simply a superposition of unresolved faint
X-ray point sources or the observed X-rays can be diffuse in origin.
We discuss each of these possibilities in turn.

\subsection{Unresolved point sources}

Various classes of faint X-rays objects have been identified in globular
clusters, including quiescent low mass X-ray binaries (qLMXBs), MSPs,
cataclysmic variables (CVs), and the chromospherically active main sequence
binaries (ABs). Although the individual contribution of these classes of systems
at the low limiting luminosities ($< L_{\rm limit}$) of our study are unknown,
we examine the possibility that the detected residual X-ray emission in the cluster
cores is a result of a blend of faint X-ray sources following the method used by
Okada et al. (2007). Specifically, we assume a cumulative luminosity function of
the form $N(>L_{X})=N_{0}L_{X}^{-q}$ and its extrapolation to luminosities,
$L_{X}$, below $L_{\rm limit}$.  The normalization $N_{0}$ can be determined from the
local source number density of X-ray sources above $L_{\rm limit}$ in each cluster core.
The slope $q$ has been inferred to be 0.5 in both NGC~6440 and NGC~6752 by Pooley et al.
(2002), but has not been determined for M28 and M62. Here, we assume it to be 0.5 in
these two clusters as well.

With the aforementioned assumptions, the contribution from the faint X-ray source
population below the detection threshold is estimated to be $4.1\times10^{31}$ erg~s$^{-1}$,
$3.1\times10^{32}$ erg~s$^{-1}$, $8.5\times10^{31}$ erg~s$^{-1}$, and $1.0\times10^{31}$
erg~s$^{-1}$ in the cores of M28, NGC~6440, M62 and NGC~6752 respectively. The estimates 
for NGC~6440 and NGC~6752 are within the $1\sigma$ uncertainities and for M62 within the $2 
\sigma$ uncertainty of the corresponding
observed unresolved X-ray luminosities. On the other hand, the contribution from unresolved
sources in M28 is about $\sim5$ times smaller than the observed value.  We note, 
however, that this preliminary result is sensitive to the adopted slope of the luminosity 
function.  For example, for a slope as steep as that inferred in the case of 47~Tuc (i.e. 
$q=0.78$ Pooley et al.  2002), the total luminosity of the unresolved point source population 
in M28 can attain a level ($\sim 1.5\times10^{32}$~s$^{-1}$) comparable to its observed value.  
Adopting such a slope for M62 similarly leads to a higher contribution corresponding to 
unresolved X-ray luminosity of $3.0\times10^{32}$~s$^{-1}$, which is close to  
the $1 \sigma$ uncertainty of the observed value. Hence, notwithstanding the spectral 
properties of the unresolved emission, which depend on the relative contribution of each 
sub population, the luminosity level of the residual X-ray emission found in the cores of 
these globular could reflect the contribution of unresolved faint sources, subject to  
our choice for extrapolation of the luminosity function to luminosities less than 
$L_{\rm limit}$.

\subsection{Diffuse emission}

In view of the uncertainties in the estimates of the contribution of a faint point
source population to the unresolved X-ray emission, we also consider the possibility that
the emission is diffuse in origin.  In this case, we explore the hypothesis that the
high energy emission with photon energy higher than several keV results from PWNe of
MSPs. The outflow accompanying their relativistic winds could be responsible for reducing the
amount of intracluster gas in the clusters (see Spergel 1991). Indeed, the low
dispersion measure for the pulsars in 47 Tuc (Freire et al. 2001) suggests that some mechanism
operates to reduce the mass of gas in the central region expected to be accumulated in
the $\sim10^{7}-10^{8}$ year interval between passages of the cluster through the Galactic
disk (cf. Camilo \& Rasio 2005).

It is known that globular clusters of high central densities host a large number of
MSPs (e.g. 47 Tuc, Terzan 5). Based on the results of \S 2, the limiting luminosities for
point source detection in the four globular clusters with sufficient photons for spectral 
analysis are in the range $1.1-18 \times 10^{30}$ergs s$^{-1}$, which overlaps with the 
range of luminosities found for most MSPs in 47 Tuc ($\sim 1-4 \times 10^{30}$ ergs s$^{-1}$; 
Grindlay et al. 2002). In addition, the spectra of MSPs in 47 Tuc are mainly thermal with 
a black body temperature $kT \sim 0.2$~keV. If we assume that the black body luminosity
component obtained by fitting the model of PL+BB listed in Table 3 results from
unresolved MSPs and the limiting luminosity listed in Table 2 is the upper limit of the
characteristic luminosity of the unresolved MSPs in various globular clusters,
the possible number of faint MSPs in the core regions of M28, NGC 6440, M62 and 
NGC 6752 could be 6, 2, 12 and 2 respectively.  

It is of some interest to compare these estimates with the uncovered pulsar population 
in each cluster core. In the case of M28, there are 6 MSPs located within one core radius 
after the exclusion of M28A (cf. Fig.~\ref{m28_coord}).  For NGC~6440 and NGC~6752, the 
radio timing positions of MSPs NGC~6440B and NGC~6752D are found to coincide with the X-ray 
sources in their cores (cf. Hui et al. 2009).  Therefore, their X-ray contribution 
has already been removed in the procedure of PSF subtraction. After excluding these two MSPs, there 
are 2 MSPs in each of these two cluster cores. Hence, the estimated numbers of MSPs based on the 
black body components in our PL+BB fittings are comparable with the known MSP population 
in the cores of these globular clusters.  

In M62, 6 MSPs have so far been discovered (D'Amico et al. 2001; Possenti et al. 2003; Chandler 
2003).  However, timing solutions have only been reported for M62A, M62B and M62C (See Table~1 in 
Possenti et al.~2003). M62B and M62C are located within one core radius. Since both have 
identified X-ray counterparts (cf. Hui et al. 2009), their contributions have been removed. 
Hence, our simple estimate suggests that a potentially additional population of MSPs 
in the core of M62 may not have been revealed in a previous radio survey.  This inference is 
not unreasonable as the stellar encounter rate in its core is comparable with that of 47~Tuc 
(cf. Fregeau 2008). On the other hand, the black body component of the unresolved emission 
can also result from other source classes, such as ABs. 

If MSPs in the core contribute to the black body component in the PL+BB fittings, the 
power-law component may arise from PWNe in which non-thermal emission emanates from the
shock region formed by the interaction of the pulsar wind with the interstellar medium
(e.g. Chevalier 2000; Cheng, Taam and Wang 2004, 2006). 
Adopting a characteristic spin-down power of $10^{34}$ ergs s$^{-1}$ for the unresolved MSPs, 
the non-thermal X-ray luminosity contributed by each PWN 
is estimated to be $\sim 10^{31}$ergs s$^{-1}$ (see Cheng, Taam 
\& Wang 2004).  For the estimated number of unresolved MSPs from the thermal component (see above), 
the total non-thermal luminosities from M28, NGC 6440, M62 and NGC 6752 are $6 \times 10^{31}$ergs 
s$^{-1}$, $2 \times 10^{31}$ ergs s$^{-1}$, $1.2 \times 10^{32}$ ergs s$^{-1}$, and $2 \times 
10^{31}$ ergs s$^{-1}$, respectively. In comparing these model predicted luminosities with the 
observed luminosities in Table 3, we find that the model predicted non-thermal emission from PWNe 
is substantially less than the observed diffuse power-law component in M28 and NGC 6440. This 
implies that the power-law component in these two globular clusters are likely contributed by 
other subpopulations (e.g., CVs and ABs). On the other hand, the model predicted non-thermal 
luminosities of M62 and NGC 6752 are comparable to the luminosities in the power-law component in 
these two globular clusters, suggesting that in this interpretation MSPs may contribute 
in these two globular clusters. However, the fitted PL index obtained in M62 
is near unity, which is a difficulty for the interpretation since the PL index 
resulting from the PWN emission should exceed 1.5. 
Taking into account the case of NGC 6752, where the black body luminosity is only an upper 
limit, we conclude that the evidence for non-thermal emission from PWNe in globular clusters 
is lacking.

We have further examined this inference by investigating the vicinity of M28A, which is the 
most powerful one among all the MSPs residing in globular clusters. The emission of 
PWN mainly arises from a region near the shock radius. Therefore, we calculate the shock 
radius of M28A by $R_s = ({\dot E}/2\pi \rho v_p^2 c)^{1/2}\sim5\times 10^{16}
{\dot E}_{34}^{1/2}n_{0.1}^{-1/2}v_{p,2}^{-1}$~cm, 
where $v_{p,2}$ is the pulsar velocity in units of 100 km s$^{-1}$, $\dot E_{34}$ is the
characteristic spin-down power in units of $10^{34}$ ergs s$^{-1}$, and $n_{0.1}$ is the
number density of the intracluster medium in units of 0.1 cm$^{-3}$ (see Cheng, Taam \& Wang 
2004).  The number density in the core of M28 is estimated as 
$n\sim\left<DM\right>/\left<d_\perp\right>\sim3.6$~cm$^{-3}$,
where $\left<DM\right>$ and $\left<d_\perp\right>$ are the standard deviations of the dispersion 
measure and the offset of the MSPs in the cluster core (cf. B\'egin 2006). 
Adopting $v_p$ to be the velocity dispersion, $\sqrt{3}\sigma_{v}=28$~km/s at the cluster center 
(where $\sigma_{v}$ is the one-dimensional velocity dispersion at the center of M28 
\footnote{http://www-int.stsci.edu/$\sim$ognedin/gc/vesc.dat}), the spin-down power of M28A implies 
the shock radius to be $\sim5.4$ arcsec at 5.5 kpc. This is larger than the region adopted 
in subtracting the point source contribution at the position of M28A. However, we do not 
identify any feature around this subtraction region in examining the residual emission (cf. 
Fig.~\ref{m28_coord}).  Therefore, the absence of any extended structure around the powerful 
MSP M28A provides further support that there is no compelling evidence for PWNe in 
globular clusters. 

\section{DISCUSSION \label{discussion}}

An observational investigation of the unresolved X-ray emission component
in the cores of globular clusters that host MSPs has been
carried out. Our sample contains only globular clusters targeted by the
Chandra X-ray Observatory in order to take advantage of its sub-arcsecond
resolution which is necessary for accurate extraction of the point source
contributions to the X-ray emission.  The subtraction of known as well as
newly discovered X-ray point sources was performed to estimate the luminosity
and spectral properties of the unresolved emission. Of the ten globular
clusters selected (whose cores are sufficiently large), but excluding 47 Tuc
and Terzan 5, detectable unresolved X-ray emission was found for M28, NGC 6440,
M62, and NGC 6752. The luminosities in the 0.3-8 keV energy band ranged by about
an order of magnitude from $\sim 1.5 \times 10^{31}$ ergs s$^{-1}$ to $\sim 2.5
\times 10^{32}$ ergs s$^{-1}$. Among these clusters, the limiting luminosity for
the point X-ray source detections is as low as $1.1 \times 10^{30}$ ergs s$^{-1}$
for NGC 6752. The spectra of the unresolved emission were found to be equally well fit,
which varied from cluster to cluster,
by a power-law characterized by a photon index, $\Gamma$, $\sim 1.5-2.3$, by a
thermal bremsstrahlung model with temperature of $\sim 2-9$ keV, or a black body of
0.18 keV plus power-law component ($\Gamma \sim 1-2$).

To assess the contribution to the X-ray luminosity in the 0.3-8 keV energy band
by faint point sources in these cluster cores, estimates were based on an
extrapolation of the luminosity function of detected pointed X-ray sources. We note
that these contributions represent a heterogenous stellar population consisting of
ABs, CVs, and MSPs. Inherent uncertainties exist in such a procedure since other 
source populations may contribute at lower luminosities that do not significantly 
contribute at higher luminosities. Given these caveats, it has been shown that the 
contribution of a faint source population (with luminosities less than the detection 
threshold for point sources) can be comparable to the unresolved X-ray emission 
measured in the cores of NGC 6440, M62, and NGC 6752.  On the other hand, the contribution 
of faint sources to the unresolved emission in M28 can be $\sim 5$ times less than the 
measured value, although taking the steeper slope for the X-ray luminosity function of 47 
Tuc leads to a faint source contribution comparable to that measured for the unresolved 
X-ray emission.

Although luminous X-ray PWNe have been detected from pulsars in the field, we do not 
find any evidence of a PWNe contribution to the diffuse X-ray emission in the globular 
clusters studied in this paper. This may suggest that the X-ray emission from PWNe is in slow 
cooling region, which has a much lower efficiency.  However, we have concluded that 
the diffuse X-ray emission results from the unresolved faint point source population in 
the cluster cores. In the following, we briefly discuss the characteristics of the 
faint source population as constrained by the properties of the unresolved X-ray emission.  

A number of possible candidate sources could contribute at the faint luminosity levels.
Among them are white dwarfs, either magnetic or non-magnetic.
For the non-magnetic white dwarfs, X-ray emission could be associated with a boundary
layer between the inner edge of the accretion disk and the white dwarf surface.
Based on the work by Pandel et al. (2005) of low accretion rate CVs,
these sources should be detectable as resolved point sources ($>10^{31}$ ergs
s$^{-1}$ corresponding to accretion rates $> 10^{-12} \msun$ yr$^{-1}$). To contribute
to the unresolved source population, one requires even lower accretion rates. 
In contrast, it is more likely that a sub population of magnetic white dwarfs 
may contribute. For example, low accretion rate polars corresponding to magnetic white 
dwarfs in detached systems (which capture a substantial fraction of the stellar wind 
of its detached companion) may be a viable population each contributing at a level of 
$\sim 10^{30}$ ergs s$^{-1}$ (see Webbink \& Wickramasinghe 2005). On the other hand, 
the white dwarfs in binaries known as polars or intermediate polars, where the mass 
transfer is accomplished via Roche lobe overflow, which are generally characterized 
by a hard spectrum with a photon indices near unity (see Heinke et al. 2008), are 
sufficiently luminous ($>10^{31}$ ergs/s) that they would have been detected as resolved 
point sources. 

Coronal emission from single stars or ABs is another possible contributor
to the faint X-ray emission; however, these sources are generally soft.  Harder spectra,
corresponding to higher coronal temperatures, are observed from such a population, but
at luminosities such that they would be individually detectable (see G\"udel 2004).
Although strong flares can produce hard spectra, it is difficult to quantify their
contribution since the time averaged luminosities of such sources is not well 
characterized.

\begin{figure}
\centerline{\psfig{figure=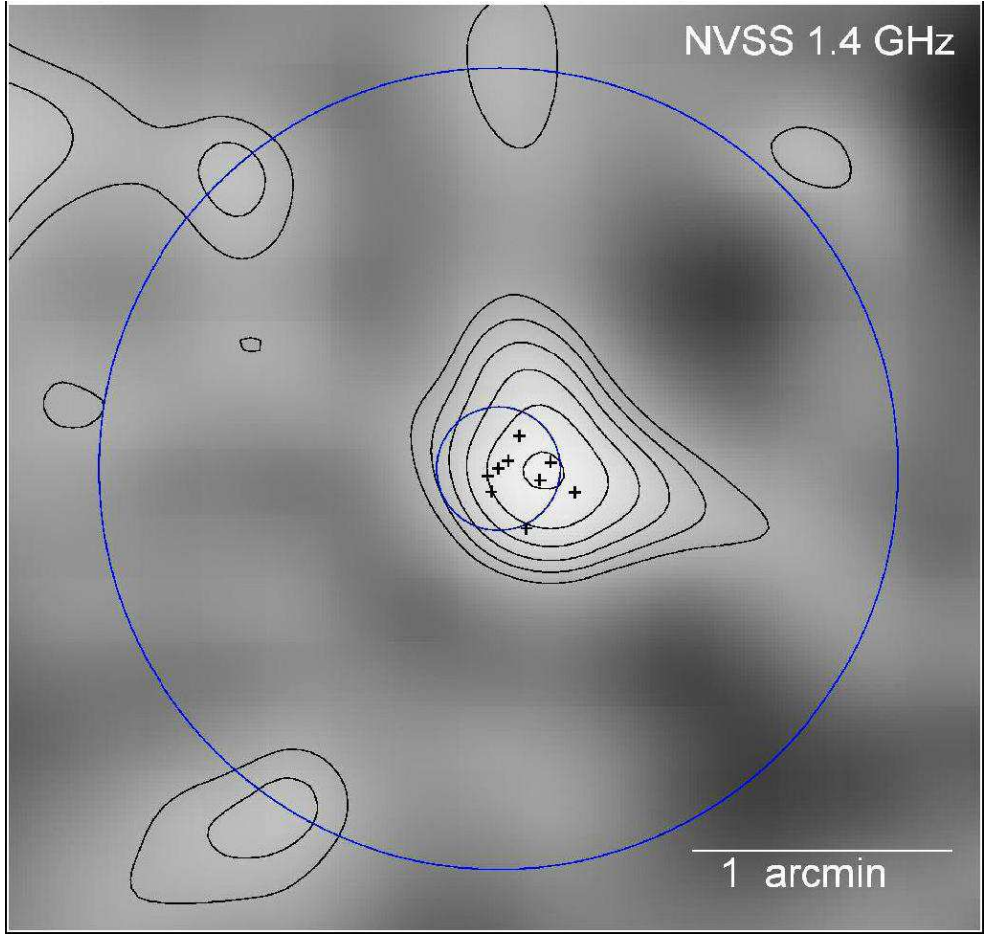,width=13cm}}
\caption{
The 1.4 GHz NVSS image of the field around M28. Top is north and left is east.
The radio pulsar positions are indicated by
black crosses. The radio contours are at the levels between 0.5 mJy/beam$-$2 mJy/beam.
The blue circles are centered at the optical center of the cluster with the radii equal to
the core radius (small) and the half-mass radius (large).
 }
\label{m28_nvss}
\end{figure}

Independent estimates of a possible undetected MSP population can be provided 
by radio data. We have examined the NRAO/VLA Sky Survey image of the field 
around M28 and found that there is an excess near the center of M28 (see 
Figure~\ref{m28_nvss}). Here, the overlaid contours were computed at the 
levels between 0.5 mJy/beam$-$2 mJy/beam. The radio timing positions of all the 
pulsars within the half-mass radius are illustrated for reference in 
Figure~\ref{m28_nvss} as crosses, which are all found to have a good correspondence 
with the radio feature. The FWHM of the feature is $\sim40$ arcsec is comparable with 
the angular resolution of NVSS data (cf. Condon et al. 1998). Assuming a beam size 
(FWHM) of 45 arcsec, the flux density of this feature is found to be $\sim2.2$ mJy. 
This compares to the sum of the flux densities of the known MSPs $\sim1.5$ mJy 
(B\'egin 2006).  Since the rms brightness fluctuation of the NVSS image is at the 
level of $\sim0.45$ mJy/beam (Condon et al. 1998), we cannot determine the possible 
contribution from the undetected MSPs on the basis of the NVSS data.  A sensitive 
radio inteferometric observation (e.g. ATCA), which can attain a sensitivity limit at 
the order of $\sim0.01$~mJy/beam, is highly desirable to provide a tight independent 
constraint on the MSP population in this cluster.

We have also searched for radio emission from NGC~6440, M62 and NGC~6752
from the NVSS data and the Sydney University Molonglo Sky Survey (SUMSS)
data (Bock et al. 1999), but did not identify any excess near their cores. 

Future observations of the core regions of globular cluster systems
at fainter limiting X-ray luminosities will be necessary to quantify
the luminosity function at lower levels to help distinguish the
various subpopulation contributions to the unresolved emission.
If additional millisecond pulsars are to contribute to the X-ray 
emission, improved determinations of the radio
timing positions of known MSPs are essential. Such a
multi-wavelength observational program is of paramount importance in
further constraining the nature and evolutionary history of the
faint X-ray point source population in these dense stellar systems.

\acknowledgments CYH and KSC were supported by the Croucher
Foundation Postdoctoral Fellowship and a GRF grant of Hong Kong
Government under HKU700908P respectively. RET was supported in part
by the Theoretical Institute for Advanced Research in Astrophysics
(TIARA) operated under the Academia Sinica Institute of Astronomy \&
Astrophysics in Taipei, Taiwan.

\clearpage

\begin{deluxetable}{lcccc}
\tablewidth{0pc}
\tablecaption{Details of the Chandra observations of the selected globular clusters.}
\startdata
\hline\hline
Obs. ID. & Detector  & Start Date \& Time & Mode & Effective Exposure \\
{} & {}  & (UTC) &  & sec \\\hline

\multicolumn{5}{c}{\bf M28 (NGC~6626)}\\
\hline
2683 & ACIS-S & 2002-09-09T16:55:03 & VFAINT & 14110  \\
2684 & ACIS-S & 2002-07-04T18:02:19 & VFAINT & 12746  \\
2685 & ACIS-S & 2002-08-04T23:46:25 & VFAINT & 13511  \\
\hline\hline
\multicolumn{5}{c}{\bf NGC~6440}\\
\hline
947 & ACIS-S & 2000-07-04T13:28:39 & FAINT &  23279 \\
3360 & ACIS-S & 2001-08-18T20:04:32  & FAINT & 2511  \\
3799 & ACIS-S & 2003-06-27T08:57:31 & FAINT & 24046  \\
\hline\hline
\multicolumn{5}{c}{\bf M62 (NGC~6266)}\\
\hline
2677 & ACIS-S & 2002-05-12T09:12:42  & FAINT & 62266  \\
\hline\hline
\multicolumn{5}{c}{\bf NGC~6752 }\\
\hline
948 & ACIS-S & 2000-05-15T04:36:02 & FAINT & 29468  \\
6612 & ACIS-S & 2006-02-10T22:48:48 & VFAINT & 37967  \\
\hline\hline
\multicolumn{5}{c}{\bf M5 (NGC~5904) }\\
\hline
2676 & ACIS-S & 2002-09-24T06:51:22 & FAINT & 44656  \\
\hline\hline
\multicolumn{5}{c}{\bf M13 (NGC~6205) }\\
\hline
5436 & ACIS-S & 2006-03-11T06:19:34 & FAINT & 26799  \\
7290 & ACIS-S & 2006-03-09T23:01:13 & FAINT & 27895  \\
\hline\hline
\multicolumn{5}{c}{\bf M3 (NGC~5272) }\\
\hline
4542 & ACIS-S & 2003-11-11T16:33:18 & VFAINT & 9932  \\
4543 & ACIS-S & 2004-05-09T17:26:32 & VFAINT & 10152  \\
4544 & ACIS-S & 2005-01-10T08:54:31 & VFAINT & 9441  \\
\hline\hline
\multicolumn{5}{c}{\bf M71 (NGC~6838) }\\
\hline
5434 & ACIS-S & 2004-12-20T15:18:45 & VFAINT & 52446  \\
\hline\hline
\multicolumn{5}{c}{\bf M53 (NGC~5024) }\\
\hline
6560 & ACIS-S & 2006-11-13T18:32:22 & VFAINT & 24565  \\
\hline\hline
\multicolumn{5}{c}{\bf M4 (NGC~6121) }\\
\hline
946 & ACIS-S & 2000-06-30 04:24:23 & VFAINT & 25816  \\
\enddata
\label{obs_info}
\end{deluxetable}

\clearpage

\begin{deluxetable}{lccccc}
\tablewidth{0pc}
\tablecaption{Properties of 10 selected globular clusters.}
\startdata
\hline\hline
Cluster Name & $\rho_{0}$\tablenotemark{a} & $E(B-V)$\tablenotemark{b} & distance & core radius & $L_{\rm limit}$\tablenotemark{c} \\
{}  & $\log$ $L_{\odot}$pc$^{-3}$ & {} & kpc & pc & erg s$^{-1}$ \\\hline\hline
M28 & 4.73 & 0.40 & 5.5 & 0.38 & $3.5\times10^{30}$ \\
NGC~6440  & 5.28 & 1.07 & 8.3 & 0.31 & $1.8\times10^{31}$ \\
M62  & 5.14 & 0.47 & 6.9 & 0.36 & $7.1\times10^{30}$ \\
NGC~6752 & 4.91 & 0.04 & 4.0 & 0.20 & $1.1\times10^{30}$ \\
M5       & 3.91 & 0.03 & 7.5 & 0.92 & $6.7\times10^{30}$ \\
M13      & 3.33 & 0.02 & 7.7 & 1.75 & $9.2\times10^{30}$ \\
M3       & 3.51 & 0.01 & 10.4 & 1.66 & $1.4\times10^{31}$ \\
M71      & 3.04 & 0.25 & 4.0 & 0.73 & $1.6\times10^{30}$ \\
M53      & 3.05 & 0.02 & 17.8 & 1.86 & $1.3\times10^{32}$ \\
M4       & 3.82 & 0.36 & 2.2 & 0.53 & $8.5\times10^{29}$ \\
\enddata
\tablenotetext{a}{Logarithm of central luminosity density}
\tablenotetext{b}{Average foreground reddening}
\tablenotetext{c}{Limiting luminosities inferred from the X-ray source detections}
\label{gc_info}
\end{deluxetable}

\clearpage

\begin{table}
\caption{X-ray spectral properties of the unresolved X-ray emission
in the globular cluster cores.}
\vspace{1cm}
\resizebox{!}{8cm}{
\begin{tabular}{lccccccccc}
\hline\hline\\
Model $^{a}$ & $\chi^{2}_{\nu}$ & D.O.F. & $n_{H}^{b}$ & $\Gamma^{c}$ & $kT_{\rm BREMSS}^{d}$ & $kT_{\rm BB}^{d}$ &
$L_{\rm PL}^{e}$ & $L_{\rm BREMSS}^{e}$ & $L_{\rm BB}^{e}$ \\
      &                  &        & $10^{21}$ cm$^{-2}$ &   & keV  & keV  &
$10^{32}$ erg~s$^{-1}$ & $10^{32}$ erg~s$^{-1}$ & $10^{31}$ erg~s$^{-1}$ \\
 \hline
 \multicolumn{10}{c}{\LARGE\bf M28}\\
 \hline\\
 PL    & 0.83 & 25 & $2.75^{+0.78}_{-0.84}$  & $2.35^{+0.41}_{-0.25}$ & - & - & $2.54^{+1.27}_{-0.62}$ & - & - \\
 \\
 BREMSS & 0.82  & 25 & $1.50^{+0.90}_{-0.76}$ & - & $2.58^{+1.09}_{-0.75}$ & - & - & $1.67^{+0.99}_{-0.62}$ & - \\
 \\
  PL+BB  & 0.84 & 25 & 2.2 (fixed) & $2.21^{+0.18}_{-0.23}$ & - & 0.18 (fixed) & $2.28^{+0.44}_{-0.56}$ & - & $<2.21$ \\
 \\
 BREMSS+BB  & 0.82 & 25 & 2.2 (fixed) & - & $2.68^{+0.87}_{-0.84}$ & 0.18 (fixed) & -  & $1.66^{+0.95}_{-0.67}$ & $2.21^{+2.19}_{-2.21}$ \\
 \\
 \hline
 \multicolumn{10}{c}{\LARGE\bf NGC 6440}\\
 \hline\\
 PL  & 0.39 & 4 & 5.9 (fixed) & $2.04^{+0.50}_{-0.46}$ & - & - & $2.00^{+1.71}_{-0.76}$ & - & - \\
 \\
 BREMSS & 0.44  & 4 & 5.9 (fixed) & -  & $3.06^{+6.18}_{-1.25}$ & - & - & $1.57^{+1.51}_{-0.74}$ & - \\
 \\
 PL+BB   & 0.51 & 3 & 5.9 (fixed) & $1.83^{+0.69}_{-0.56}$ & - & 0.18 (fixed) & $1.77^{+3.02}_{-1.33}$ & - & $3.61^{+11.49}_{-3.61}$ \\
 \\
 BREMSS+BB  & 0.44 & 4 & 5.9 (fixed) & - & 3.06 (fixed) & 0.18 (fixed) & - & $1.54^{+0.27}_{-0.33}$ & $1.19^{+7.49}_{-1.19}$ \\
 \\
 \hline
\multicolumn{10}{c}{\LARGE\bf M62}\\
 \hline\\
 PL    & 0.43 & 8 & $<0.52$  &  $1.54^{+0.22}_{-0.16}$ & - & - & $1.50^{+0.62}_{-0.36}$ & - & - \\
 \\
 BREMSS & 0.58 & 8 & $<0.35$ & - & $9.23^{+17.50}_{-4.10}$ & - & - & $1.41^{+0.26}_{-0.27}$ & - \\
 \\
 PL+BB & 0.34 & 8 & 2.6 (fixed) & $1.03^{+0.17}_{-0.18}$ & - & 0.18 (fixed) & $1.50^{+1.46}_{-0.74}$ & - & $8.19^{+1.97}_{-2.45}$ \\
 \\
 BREMSS+BB & 0.53 & 9 & 2.6 (fixed) & - & 9.23 (fixed) & 0.18 (fixed) & - & $1.33\pm0.19$ & $6.15\pm1.58$ \\
 \\
 \hline
 \multicolumn{10}{c}{\LARGE\bf  NGC 6752}\\
 \hline\\
 PL    & 0.95 & 7 & 0.22 (fixed) & $1.94^{+0.39}_{-0.36}$ & - & - & $0.15^{+0.08}_{-0.04}$ & - & - \\
 \\
 BREMSS  & 0.89  & 7 & 0.22 (fixed) & - & $2.38^{+3.06}_{-1.00}$ & - & - & $0.12^{+0.11}_{-0.06}$ & - \\
 \\
 PL+BB  & 0.95 & 7 & 0.22 (fixed) & 1.94 (fixed) & - & 0.18 (fixed) & $0.15^{+0.02}_{-0.04}$ & - & $<0.23$ \\
 \\
 BREMSS+BB  & 1.04 & 6 & 0.22 (fixed) & - & $2.38^{+3.08}_{-1.00}$ & 0.18 (fixed) & - & $0.12^{+0.11}_{-0.07}$ & $<0.33$ \\
 \\
\hline\hline
 \end{tabular}
              }
\vspace{0.5cm}

$^{a}$ PL=Power-law; BREMSS=Thermal bremsstrahlung; BB=Blackbody \\
$^{b}$ Hydrogen column density\\
$^{c}$ X-ray photon index\\
$^{d}$ $kT_{\rm BB}$/$kT_{\rm BREMSS}$ is the temperature of the black body/thermal bremsstrahlung model.\\
$^{e}$ $L_{\rm PL}$/$L_{\rm BREMSS}$/$L_{\rm BB}$ is the absorption-corrected luminosity inferred from the
power-law/thermal bremsstrahlung/black body component.\\
\label{gc_spec}
\end{table}

\clearpage

\clearpage

\clearpage

\clearpage

\end{document}